\documentclass[english]{article}
\usepackage[T1]{fontenc}
\usepackage[latin9]{inputenc}
\usepackage[a4paper]{geometry}
\usepackage{textcomp}
\usepackage{amsmath}
\usepackage{setspace}
\usepackage{amssymb}
\makeatletter
\newcommand{\lyxaddress}[1]{
\par {\raggedright #1
\vspace{1.4em}
\noindent\par}
}

\makeatother

\usepackage{babel}

\begin{document}

\title{Gauge Formulation for Two Potential Theory of Dyons }

\author{O. P. S. Negi%
\thanks{Permanent Address:- \textbf{Department of Physics, Kumaun University,
S. S. J. Campus, Almora-263601(Uttarakhand), India}%
} ~and H. Dehnen }

\maketitle

\lyxaddress{\begin{center}
Universität Konstanz\\
 Fachbereich Physik\\
 Postfach-M 677\\
 D-78457 Konstanz, Germany
\par\end{center}}

\lyxaddress{\begin{center}
Email:- ops\_negi@yahoo.co.in\\
 Heinz.Dehnen@uni-konstanz.de
\par\end{center}}
\begin{abstract}
Dual electrodynamics and corresponding Maxwell's equations ( in the
presence of monopole only) are revisited from the symmetry of duality
and gauge invariance. Accordingly, the manifestly covariant, dual
symmetric and gauge invariant two potential theory of generalized
electromagnetic fields of dyons has been developed consistently from
$U(1)\times U(1)$ gauge symmetry. Corresponding field equations and
equation of motion are derived from Lagrangian formulation adopted
for $U(1)\times U(1)$ gauge symmetry for the justification of two
four potentials of dyons.

\textbf{Key words}- Dual electrodynamics, duality, gauge invariance,
monopoles and dyons

\textbf{PACS No.}- 14.80 Hv
\end{abstract}

\section{Introduction}

		The asymmetry between electricity and magnetism has become clear
at the end of 19th century with the formulation of Maxwell's equations.
and physicists were fascinated about the idea of magnetic monopoles.
Dirac \cite{key-1} put forward the idea of magnetic monopoles to
symmetrize Maxwell's equations and showed that the quantum mechanics
of an electrically charged particle of charge $\mathsf{e}$ and a
magnetically charged particle of charge $\mathsf{g}$ is consistent
only if $\mathsf{eg}=2\pi\, n$, $n$ being an integer. Schwinger-Zwanziger
\cite{key-2} generalized this condition to allow for the possibility
of particles (dyons) that carry both electric and magnetic charge.
A quantum mechanical theory can have two particles of electric and
magnetic charges $(\mathsf{e_{1},\, g_{1}})$ and $(\mathsf{e_{2},\, g_{2}})$
only if $\mathsf{e_{1}}\mathsf{\mathsf{g}}_{2}-\mathsf{e_{2}}\mathsf{g}_{1}=2\pi\, n$.
Fresh interests in this subject have been enhanced by 't Hooft -Polyakov
\cite{key-3} with the idea that the classical solutions having the
properties of magnetic monopoles may be found in Yang - Mills gauge
theories. Julia and Zee \cite{key-4} extended the 't Hooft-Polyakov
theory \cite{key-3} of monopoles and constructed the theory of non
Abelian dyons. It is now widely recognized that the standard model,
which combines the gauge theory of strong interactions with the model
of electro weak interaction, is a gauge theory that contains monopole
and dyon solutions. The quantum mechanical excitation of fundamental
monopoles include dyons which are automatically arisen \cite{key-5}
from the semi-classical quantization of global charge rotation degree
of freedom of monopoles. In view of the explanation of CP-violation
in terms of non-zero vacuum angle of world \cite{key-6}, the monopoles
are necessary dyons and Dirac quantization condition permits dyons
to have analogous electric charge. Renewed interests in the subject
of monopole has gathered \cite{key-7,key-8,key-9,key-10} enormous
potential importance in connection current grand unified theories,
supersymmetric gauge theories and super strings. But unfortunately
the experimental searches \cite{key-11} for these elusive particles
have proved fruitless as the monopoles are expected to be super heavy
and their typically masses are about two orders of magnitude heavier
than the super heavy $X$ bosons mediating proton decay. However,
a group of physicists \cite{key-12} are now claiming that they have
found indirect evidences for monopoles and now it is being speculated
that magnetic monopoles may play an important role in condensed matter
physics. In spite of the enormous potential importance of monopoles
(dyons) and the fact that these particles have been extensively studied,
there has been presented no reliable theory which is as conceptually
transparent and predictably tact-able as the usual electrodynamics
and the formalism necessary to describe them has been clumsy and not
manifestly covariant. On the other hand, the concept of electromagnetic
(EM) duality has been receiving much attention \cite{key-9,key-10}
in gauge theories, field theories, supersymmetry and super strings.
So, keeping in view the recent potential importance of monopoles (dyons)
and the applications of electromagnetic duality, in this paper, we
have made an attempt to revisit the analogous consistent formulation
of dual electrodynamics subjected by the magnetic monopole only. Gauge
formulation has been adopted accordingly to derive the dual Maxwell's
equation, equation of motion and Bianchi identity for dual electric
charge (i.e. magnetic monopole) from the minimum action principle.
Accordingly, we have discussed the dual symmetric and manifestly covariant
formulation of generalized fields of dyons in order to obtain the
generalized Dirac-Maxwell's (GDM) field equations and Lorentz force
equation of motion of dyons in terms of two four potentials. Two potential
theory of magnetic monopoles (dyons) have been justified from $U(1)\times U(1)$
gauge symmetry. Consequently, the gauge symmetric and dual invariant
manifestly covariant theory has been reformulated consistently from
the $U(1)\times U(1)$ gauge symmetry. It has been emphasized that
the two $U(1)$ gauge group acts in different manner whereas the first
$U^{(e)}(1)$ acts on the Dirac spinors while the other group $U^{(m)}(1)$
acts on Dirac iso-spinors. We have also developed accordingly the
consistent Lagrangian formulation for the justification of two gauge
potentials of dyons.

\section{Dual Electrodynamics}

	 Duality invariance is an old idea introduced a century ago in classical
electromagnetism \cite{key-10} for the following Maxwell's equations
in vacuum i.e.

\begin{align}
\overrightarrow{\nabla}\cdot\overrightarrow{E} & =0\nonumber \\
\overrightarrow{\nabla}\cdot\overrightarrow{B}= & 0\nonumber \\
\overrightarrow{\nabla}\times\overrightarrow{E} & =-\frac{\partial\overrightarrow{B}}{\partial t}\nonumber \\
\overrightarrow{\nabla}\times\overrightarrow{B}= & \frac{\partial\overrightarrow{E}}{\partial t}\label{eq:1}\end{align}
where $\overrightarrow{E}$ and $\overrightarrow{B}$ are respectively
the the electric and magnetic field strengths and for brevity we use
natural units $c=\hslash=1$, space-time four-vector $\left\{ x^{\mu}\right\} =(t,\, x,\, y,\, z),$$\{x_{\mu}=\eta_{\mu\nu}x^{\mu}\}$
and $\{\eta_{\mu\nu}=+1,-1,-1,-1=\eta^{\mu\nu}\}$ through out the
text. Maxwell's equations (\ref{eq:1}) are invariant not only under
Lorentz and conformal transformations but are also invariant under
the following duality transformations,

\begin{align}
\overrightarrow{E} & \Longrightarrow\overrightarrow{E}\,\cos\vartheta+\overrightarrow{B}\,\sin\vartheta\nonumber \\
\overrightarrow{B} & \Longrightarrow-\overrightarrow{E}\,\sin\vartheta+\overrightarrow{B}\,\cos\vartheta\label{eq:2}\end{align}
where $\overrightarrow{E}$ and and $\overrightarrow{B}$ are respectively
the the electric and magnetic field strengths. For a particular value
of $\vartheta=\frac{\pi}{2}$, equations (\ref{eq:2}) reduces to

\begin{align}
\overrightarrow{E} & \longmapsto\overrightarrow{B};\,\,\,\,\,\,\,\overrightarrow{B}\longmapsto-\overrightarrow{E}\label{eq:3}\end{align}
which can be written as 

\begin{align}
\left(\begin{array}{c}
\overrightarrow{E}\\
\overrightarrow{B}\end{array}\right) & \Longrightarrow\left(\begin{array}{cc}
0 & 1\\
-1 & 0\end{array}\right)\left(\begin{array}{c}
\overrightarrow{E}\\
\overrightarrow{B}\end{array}\right).\label{eq:4}\end{align}
 Consequently, Maxwell's equations may be solved by introducing the
concept of vector potential in either two ways \cite{key-13}. 

\textbf{Case-I} : The conventional choice is being used as 

\begin{align}
\overrightarrow{E}=- & \frac{\partial\overrightarrow{A}}{\partial t}-\overrightarrow{\nabla}\phi\nonumber \\
\overrightarrow{B}= & \overrightarrow{\nabla}\times\overrightarrow{A}\label{eq:5}\end{align}
where $\left\{ A^{\mu}\right\} =(\phi,\,\overrightarrow{A})$ is described
as the four potential. So, the dual symmetric and Lorentz covariant
Maxwell's equations (\ref{eq:1}) are written in as

\begin{align}
\partial_{\nu}F^{\mu\nu}= & 0\nonumber \\
\partial_{\nu}\widetilde{F^{\mu\nu}}= & 0\label{eq:6}\end{align}
where $F^{\mu\nu}=\partial^{\nu}A^{\mu}-\partial^{\mu}A^{\nu}=A^{\mu,\nu}-A^{\nu,\mu}$
is anti-symmetric electromagnetic field tensor, $\widetilde{F^{\mu\nu}}=\frac{1}{2}\epsilon^{\mu\nu\lambda\omega}F_{\lambda\omega}$
$(\forall\mu,\nu,\eta,\lambda=0,1,2,3)$ is the dual of electromagnetic
field tensor and $\epsilon^{\mu\nu\lambda\omega}$ is the four index
Levi-Civita symbol. $\epsilon^{\mu\nu\lambda\omega}$=+1$\forall$
($\mu\nu\lambda\omega=0123)$ for cyclic permutation; $\epsilon^{\mu\nu\lambda\omega}=-1$
for any two permutations and $\epsilon^{\mu\nu\lambda\omega}=0$ if
any two indices are equal. Using equation (\ref{eq:5}), we may obtain
the electric and magnetic fields as the components of anti-symmetric
electromagnetic field tensors $F^{\mu\nu}$ and $\widetilde{F^{\mu\nu}}$
given by 

\begin{align}
F^{0j}=E_{j};\,\,\, & F^{jk}=\varepsilon^{jkl}B_{l}\,(\forall j,k,l=1,2,3)\nonumber \\
\widetilde{F^{0j}}=B_{j};\,\,\,\, & \widetilde{F^{jk}}=\varepsilon^{jkl}E_{l}\,(\forall j,k,l=1,2,3)\label{eq:7}\end{align}
where $\varepsilon^{jkl}$ is three index Levi-Civita symbol and $\varepsilon^{jkl}=+1$
for cyclic, $\varepsilon^{jkl}=-1$ for anti-cyclic permutations and
$\varepsilon^{jkl}=0$ for repeated indices. The duality symmetry
is lost if electric charge and current source densities enter to the
conventional inhomogeneous Maxwell's equations given by 

\begin{align}
\overrightarrow{\nabla}\cdot\overrightarrow{E} & =\rho\nonumber \\
\overrightarrow{\nabla}\cdot\overrightarrow{B}= & 0\nonumber \\
\overrightarrow{\nabla}\times\overrightarrow{E} & =-\frac{\partial\overrightarrow{B}}{\partial t}\nonumber \\
\overrightarrow{\nabla}\times\overrightarrow{B}=\overrightarrow{j}+ & \frac{\partial\overrightarrow{E}}{\partial t}\label{eq:8}\end{align}
where $\rho$ and $\overrightarrow{j}$ are described as charge and
current source densities which are the components of electric four-current
$\left\{ j^{\mu}\right\} =(\rho\,,\,\overrightarrow{j})$ source density.
So, the covariant form of Maxwell's equation (\ref{eq:8}) is described
as

\begin{align}
\partial_{\nu}F^{\mu\nu}= & j^{\mu}\nonumber \\
\partial_{\nu}\widetilde{F^{\mu\nu}}= & 0.\label{eq:9}\end{align}
Here, we may see that the pair $\overrightarrow{(\nabla}\cdot\overrightarrow{B}=0;\,\nabla\times\overrightarrow{E}=-\frac{\partial\overrightarrow{B}}{\partial t}$)
of Maxwell's equations (\ref{eq:8}) is described by $\partial_{\nu}\widetilde{F^{\mu\nu}}=0$
in equation (\ref{eq:9}). It has become kinematical while the dynamics
is contained in another pair $(\overrightarrow{\nabla}\cdot\overrightarrow{E}=\rho;\,\nabla\times\overrightarrow{B}=\overrightarrow{j}+\frac{\partial\overrightarrow{E}}{\partial t}$)
of Maxwell's equations (\ref{eq:8}) which described as $\partial_{\nu}F^{\mu\nu}=j^{\mu}$
in equation (\ref{eq:9}) and also reduces to following wave equation
in the presence of Lorentz gauge condition $\partial_{\mu}A^{\mu}=0$
i.e 

\begin{align}
\square A^{\mu}= & j^{\mu}\label{eq:10}\end{align}
where $\square=\frac{\partial^{2}}{\partial t^{2}}-\frac{\partial^{2}}{\partial x^{2}}-\frac{\partial^{2}}{\partial y^{2}}-\frac{\partial^{2}}{\partial z^{2}}$
is the D' Alembertian operator. So, a particle of mass $\mathtt{\mathsf{m}}$
, electric charge $\mathsf{\mathsf{e}}$ moving with a velocity $\left\{ u^{\nu}\right\} $
in an electromagnetic field is subjected by a Lorentz force given
by 

\begin{align}
\mathsf{m}\frac{d^{2}x_{\mu}}{d\tau^{2}}=\frac{dp_{\mu}}{d\tau}=f_{\mu} & =\mathtt{\mathsf{e}}F_{\mu\nu}u^{\nu}\label{eq:11}\end{align}
where $\{\ddot{x}_{\mu}\}\mbox{ is the four-acceleration, \ensuremath{f_{\mu}}}$
is four force and $p_{\mu}$ is four momentum of a particle. Equation
(\ref{eq:11}) is reduced to 

\begin{align}
\overrightarrow{f}=\frac{d\overrightarrow{p}}{dt}=\mathsf{m}\frac{d^{2}\overrightarrow{x}}{dt^{2}} & =\mathtt{\mathsf{e}}\left[\overrightarrow{E}+\overrightarrow{u}\times\overrightarrow{B}\right]\label{eq:12}\end{align}
where $\overrightarrow{p}$, $\overrightarrow{f}$ , $\overrightarrow{x}$
and $\overrightarrow{u}$ are respectively the three vector forms
of momentum, force, displacement and velocity of a particle. Here
we may observe that the Lorentz force equation of motion (\ref{eq:11}-\ref{eq:12})
are also not invariant under duality transformations (\ref{eq:3}-\ref{eq:4}).

\textbf{Case-II} : On the other hand, let us introduce \cite{key-13}
the another alternative way instead of equation (\ref{eq:5}) to write 

\begin{align}
\overrightarrow{B}=- & \frac{\partial\overrightarrow{C}}{\partial t}-\overrightarrow{\nabla}\varphi\nonumber \\
\overrightarrow{E}= & -\overrightarrow{\nabla}\times\overrightarrow{C}\label{eq:13}\end{align}
where a new potential $\left\{ C^{\mu}\right\} =(\varphi,\,\overrightarrow{C})$
is introduced \cite{key-13,key-14} as the alternative to $\left\{ A^{\mu}\right\} $.
Thus, we see that source free (homogeneous) Maxwell's equations are
same as those of equations (\ref{eq:1}) but the inhomogeneous Maxwell's
equations (\ref{eq:8}) are changed to

\begin{align}
\overrightarrow{\nabla}\cdot\overrightarrow{E}= & 0\nonumber \\
\overrightarrow{\nabla}\cdot\overrightarrow{B}= & \varrho\nonumber \\
\overrightarrow{\nabla}\times\overrightarrow{B}= & \frac{\partial\overrightarrow{E}}{\partial t}\nonumber \\
\nabla\times\overrightarrow{E}= & -\overrightarrow{\kappa}-\frac{\partial\overrightarrow{B}}{\partial t}\label{eq:14}\end{align}
subjected by the introduction of a new four current source density
$\left\{ k^{\mu}\right\} =(\varrho,\,\overrightarrow{\kappa})$. In
equation (\ref{eq:14}) we see that the pair ($\overrightarrow{\nabla}\cdot\overrightarrow{E}=0;\overrightarrow{\nabla}\times\overrightarrow{B}=\frac{\partial\overrightarrow{E}}{\partial t}$)
becomes kinematical while the dynamics is contained in the second
pair ($\overrightarrow{\nabla}\cdot\overrightarrow{B}=\varrho;\nabla\times\overrightarrow{E}=-\overrightarrow{\kappa}-\frac{\partial\overrightarrow{B}}{\partial t}$).
Equation (\ref{eq:14}) may also be written in following covariant
forms

\begin{align}
\partial_{\nu}F^{\mu\nu}= & 0\nonumber \\
\partial_{\nu}\widetilde{F^{\mu\nu}}= & k^{\mu}\label{eq:15}\end{align}
where $\widetilde{F^{\mu\nu}}=\partial^{\nu}C^{\mu}-\partial^{\mu}C^{\nu}$;
$\widetilde{\widetilde{F^{\mu\nu}}}=F^{\mu\nu}$; $\left\{ k^{\mu}\right\} =(\varrho,\overrightarrow{\kappa})$
and $\left\{ k_{\mu}\right\} =(\varrho,-\overrightarrow{\kappa})$.
Equation (\ref{eq:14}) may also be obtained on applying the transformations
(\ref{eq:3}) and (\ref{eq:4}) to equation (\ref{eq:8}) followed
by following duality transformations for potential, current and antisymmetric
electromagnetic field tensors as 

\begin{align}
A^{\mu} & \rightarrow C^{\mu};C^{\mu}\longmapsto-A^{\mu}\Longleftrightarrow\left(\begin{array}{c}
A^{\mu}\\
C^{\mu}\end{array}\right)\Rightarrow\left(\begin{array}{cc}
0 & 1\\
-1 & 0\end{array}\right)\left(\begin{array}{c}
A^{\mu}\\
C^{\mu}\end{array}\right)\nonumber \\
j^{\mu} & \rightarrow k^{\mu};k^{\mu}\longmapsto-j^{\mu}\Longleftrightarrow\left(\begin{array}{c}
j^{\mu}\\
k^{\mu}\end{array}\right)\Rightarrow\left(\begin{array}{cc}
0 & 1\\
-1 & 0\end{array}\right)\left(\begin{array}{c}
j^{\mu}\\
k^{\mu}\end{array}\right)\nonumber \\
F^{\mu\nu}\longmapsto & \widetilde{F^{\mu\nu}}\Longleftrightarrow\left(\begin{array}{c}
F^{\mu\nu}\\
\widetilde{F^{\mu\nu}}\end{array}\right)\Rightarrow\left(\begin{array}{cc}
0 & 1\\
-1 & 0\end{array}\right)\left(\begin{array}{c}
F^{\mu\nu}\\
\widetilde{F^{\mu\nu}}\end{array}\right).\label{eq:16}\end{align}
As such, we may identify the potential $\left\{ C^{\mu}\right\} =(\varphi,\,\overrightarrow{B})$
as the dual of potential $\left\{ A^{\mu}\right\} $ and the current
$\left\{ k^{\mu}\right\} =(\varrho,\overrightarrow{\kappa})$ as the
dual of current $\left\{ j^{\mu}\right\} $. Correspondingly, the
differential equations (\ref{eq:13}) are identified as the dual Maxwell's
equations. So, accordingly, we may develop the electrodynamics of
a charged particle with the charge dual to the electric charge (i.e
magnetic monopole). Applying the the electromagnetic duality to the
Maxwell's equations, we may establish the connection between electric
and magnetic charge (monopole) \cite{key-16,key-17}, in the same
manner as an electric charge $\mathsf{e}$ interacts with electric
field and the dual charge (magnetic monopole) $\mathsf{g}$ interacts
with magnetic field, as,

\begin{align}
\mathsf{\mathsf{e}} & \rightarrow\mathtt{\mathsf{g}};\,\,\,\mathsf{\mathtt{\mathsf{g}}}\rightarrow-\mathtt{\mathsf{e}}\Longleftrightarrow\left(\begin{array}{c}
\mathsf{e}\\
\mathsf{\mathsf{\mathtt{\mathsf{g}}}}\end{array}\right)\Rightarrow\left(\begin{array}{cc}
0 & 1\\
-1 & 0\end{array}\right)\left(\mathsf{\begin{array}{c}
\mathsf{e}\\
\mathtt{\mathsf{g}}\end{array}}\right)\label{eq:17}\end{align}
where $\mathsf{g}$ is described as the dual electric charge (charge
of magnetic monopole). Hence, we may recall the dual electrodynamics
as the dynamics of pure magnetic monopole. consequently, the corresponding
dynamical variables associated therein are described as the dynamical
variables in the theory of magnetic monopole. So, we may write the
new electromagnetic field tensor $\mathcal{F}_{\mu\nu}$ in place
of $\widetilde{F^{\mu\nu}}$ as

\begin{align}
\widetilde{F^{\mu\nu}}\longmapsto\mathcal{F}_{\mu\nu} & =\partial_{\nu}C_{\mu}-\partial_{\mu}C_{\nu}\,(\mu,\nu=1,2,3)\label{eq:18}\end{align}
which reproduces the following definition of magneto-electric fields
of monopole as

\begin{align}
\mathcal{F}_{0i} & =\mathsf{B}^{i},\nonumber \\
\mathcal{F}_{ij} & =-\varepsilon_{ijk}\mathsf{E}^{k}.\label{eq:19}\end{align}
Hence the covariant form of Maxwell's equations (\ref{eq:12}) for
magnetic monopole may now be written as

\begin{align}
\mathcal{F_{\mathit{\mu\nu,\nu}}}=\partial^{\nu}\mathsf{\mathbb{\mathcal{F_{\mu\nu}}}} & =k_{\mu},\nonumber \\
\widetilde{\mathcal{F}_{\mu\nu,\nu}} & =\partial^{\nu}\mathsf{\mathbb{\widetilde{\mathcal{F_{\mu\nu}}}}}=0\label{eq:20}\end{align}
where $\{k_{\mu}\}=\{\varrho,-\overrightarrow{\kappa}\}$ is the four
- current density due to the presence of the magnetic charge $\mathsf{g}$.
Accordingly, the wave equation (\ref{eq:15}) for pure monopole is
described as 

\begin{align}
\square C_{\mu} & =k_{\mu}\label{eq:21}\end{align}
in presence of Lorentz gauge condition $\partial_{\mu}C^{\mu}=0$.
Accordingly, we may develop the classical Lagrangian formulation in
order to obtain the field equation (dual Maxwell's equations) and
equation of motion for the dynamics of a dual charge (magnetic monopole)
interacting with electromagnetic field. So, the Lorentz force equation
of motion for a dual charge (i.e magnetic monopole) may now be written
from the duality equations (\ref{eq:3}) and (\ref{eq:4}) as 

\begin{eqnarray}
\frac{d\overrightarrow{p}}{d\tau}=\mathsf{\overrightarrow{f}}=\mathsf{m} & \overrightarrow{\ddot{x}}= & \mathsf{g}(\overrightarrow{B}\,-\,\overrightarrow{u}\times\overrightarrow{E})\label{eq:22}\end{eqnarray}
where $\overrightarrow{p}=\mathsf{m}\,\overrightarrow{\dot{x}}=\mathsf{m}\,\overrightarrow{u}$
is the momentum, and $\overrightarrow{f}$ is a force acting on a
particle of charge $\mathsf{g}$, mass $\mathsf{m}$ and moving with
the velocity $\overrightarrow{v}$ in electromagnetic fields. Equation
(\ref{eq:20}) can be generalized to write it in the following four
vector formulation as

\begin{eqnarray}
\mathsf{m}\frac{d^{2}x_{\mu}}{d\tau^{2}}=\frac{dp_{\mu}}{d\tau} & =\mathsf{f_{\mu}}=\mathsf{m}\ddot{x}_{\mu} & =g\mathcal{F_{\mu\nu}}u^{\nu}\label{eq:23}\end{eqnarray}
where $\left\{ u_{\nu}\right\} $ is the four velocity, $\left\{ p_{\mu}\right\} $
is four momentum, $\mathsf{f_{\mu}}$ is four force and $\{\ddot{x}_{\mu}\}$
is the four-acceleration of a particle carrying the dual charge (namely
magnetic monopole).

\section{Gauge Symmetry and Dual Electrodynamics}

Let us define a Dirac field $\psi$ with dual (magnetic) charge $\mathsf{g}$
for which the free Dirac Lagrangian is 

\begin{align}
\mathcal{L}_{0}= & \overline{\psi}(i\gamma^{\mu}\partial_{\mu}+\mathsf{m})\psi\,(i=\sqrt{-1})\label{eq:24}\end{align}
where $\gamma^{0}=\left[\begin{array}{cc}
1 & 0\\
0 & -1\end{array}\right]$ and $\gamma^{j}=\left[\begin{array}{cc}
0 & \tau^{j}\\
-\tau^{j} & 0\end{array}\right]$ are $4\times4$ complex Dirac matrices with $0$ , $1$ and $\tau^{j}$
are respectively the $2\times2$ null, unit and Pauli Matrices $\forall j=1,2,3$.
Also the gamma matrices satisfy the property

\begin{align}
\left\{ \gamma^{\mu},\gamma^{\nu}\right\}  & =2\eta^{\mu\nu}\label{eq:25}\end{align}
and $\overline{\psi}=\psi^{\dagger}\gamma^{0}$ with ($\dagger$)
denotes the Hermitian conjugation. So, the Dirac Lagrangian (\ref{eq:24})
is clearly invariant under the global gauge transformation

\begin{align}
\psi\longmapsto\psi'\longmapsto & \exp\{i\mathsf{g}\alpha)\}\psi\nonumber \\
\overline{\psi}\longmapsto\overline{\psi}'\longmapsto\overline{\psi} & \exp\{-i\mathsf{g}\alpha)\}\label{eq:26}\end{align}
where $\alpha$ is independent of space-time. So, like electromagnetism,
we elevate this symmetry to invariance under local gauge transformation 

\begin{align}
\psi\longmapsto\psi'\longmapsto & \exp\{i\mathsf{g}\alpha(x)\}\psi\nonumber \\
\overline{\psi}\longmapsto\overline{\psi}'\longmapsto\overline{\psi} & \exp\{-i\mathsf{g}\alpha(x)\}\label{eq:27}\end{align}
so that the Lagrangian (\ref{eq:24}) transforms as

\begin{align}
\overline{\psi}(i\gamma^{\mu}\partial_{\mu}+\mathsf{m})\psi\longmapsto & \overline{\psi}(i\gamma^{\mu}\partial_{\mu}-\mathsf{g\gamma^{\mu}\partial_{\mu}\alpha}+\mathsf{m})\psi.\label{eq:28}\end{align}
Since the extra term $\mathsf{g\gamma^{\mu}\partial_{\mu}\alpha}$
looks like a gauge transformation of potential, we may couple the
gauge field $C_{\mu}$ with $\psi$ so that the Lagrangian has the
local gauge symmetry. We, thus, write the Lagrangian (\ref{eq:24})
as 

\begin{align}
\mathcal{L}= & \overline{\psi}(i\gamma^{\mu}\nabla_{\mu}+\mathsf{m})\psi\label{eq:29}\end{align}
where the covariant derivative in equation (\ref{eq:28}) is given
by

\begin{align}
\nabla_{\mu} & =\partial_{\mu}-i\,\mathsf{g}C_{\mu}.\label{eq:30}\end{align}
Hence the Lagrangian (\ref{eq:24}) is invariant under the combined
gauge transformation 

\begin{align}
\psi(x)\longmapsto\psi'(x)\longmapsto & \exp\{i\mathsf{g}\alpha(x)\}\psi(x)\nonumber \\
\overline{\psi}(x)\longmapsto\overline{\psi}'(x)\longmapsto & \overline{\psi}(x)\exp\{-i\mathsf{g}\alpha(x)\}\nonumber \\
C_{\mu}\longmapsto C_{\mu}^{'} & =C_{\mu}+\partial_{\mu}\alpha(x)\label{eq:31}\end{align}
and the covariant derivative is transformed as 

\begin{align}
\nabla_{\mu}^{'}\psi' & \longmapsto\exp\{i\mathsf{g}\chi(x)\}(\nabla_{\mu}\psi).\label{eq:32}\end{align}
 As such, the Lagrangian for total dual quantum electrodynamics is
subjected by 

\begin{align}
\mathcal{L}=-\frac{1}{4} & \mathcal{F}_{\mu\nu}\mathcal{F^{\mu\nu}}+\overline{\psi}(i\gamma^{\mu}\partial_{\mu}+\mathsf{m})\psi-C_{\mu}k^{\mu}\label{eq:33}\end{align}
where \begin{align}
k^{\mu} & =\mathsf{g}\overline{\psi}(x)\gamma^{\mu}\psi(x).\label{eq:34}\end{align}
Accordingly, we get 

\begin{align}
\left[\nabla_{\mu},\nabla_{\nu}\right]\psi(x) & =-i\mathsf{g\mathcal{F}_{\mu\nu}\psi}(x)\label{eq:35}\end{align}
and with the use of Jacobi identity 

\begin{align}
\left[\nabla_{\mu},\left[\nabla_{\nu},\nabla_{\lambda}\right]\right]+\left[\nabla_{\nu},\left[\nabla_{\lambda},\nabla_{\mu}\right]\right]+\left[\nabla_{\lambda},\left[\nabla_{\mu},\nabla_{\nu}\right]\right] & =0\label{eq:36}\end{align}
we get the Bianchi identity

\begin{align}
\nabla_{\mu}\mathcal{F}_{\nu\lambda}+\nabla_{\nu} & \mathcal{F}_{\lambda\mu}+\nabla_{\lambda}\mathcal{F}_{\mu\nu}=0\label{eq:37}\end{align}
which is the kinematical statement equivalent to $\widetilde{\mathcal{F}_{\mu\nu,\nu}}=\partial^{\nu}\mathsf{\mathbb{\widetilde{\mathcal{F_{\mu\nu}}}}}=0$
for dual electrodynamics analogous to the kinematical statement $\partial_{\nu}\widetilde{F^{\mu\nu}}=0$
for usual electrodynamics. As such dual symmetry of electrodynamics
requires the existence of dual electric charge (i.e magnetic monopole).

\section{Dual Symmetric Covariant Formulation of Dyons}

Magnetic monopole has been introduced by Dirac\cite{key-1} in a different
manner in order to symmetrize Maxwell's equations (\ref{eq:8}) from
duality principle as 

\begin{align}
\overrightarrow{\nabla}\cdot\overrightarrow{E} & =\rho\nonumber \\
\overrightarrow{\nabla}\cdot\overrightarrow{B}= & \varrho\nonumber \\
\overrightarrow{\nabla}\times\overrightarrow{E} & =-\frac{\partial\overrightarrow{B}}{\partial t}-\overrightarrow{k};\nonumber \\
\overrightarrow{\nabla}\times\overrightarrow{B}= & \overrightarrow{j}+\frac{\partial\overrightarrow{E}}{\partial t}.\label{eq:38}\end{align}
We refer these differential equations as the generalized Dirac-Maxwell's
(GDM) equations. Since Dirac theory contains the controversial string
variables, so it has been generalized \cite{key-2,key-4,key-16,key-17}
to the theory of the particles carrying simultaneously the electric
and magnetic charges namely dyons. As such, GDM equations (\ref{eq:38})
are described as the field equations of dyons. The GDM equations (\ref{eq:38})
may directly be obtained on combining the Maxwell's equations (\ref{eq:8})
and their dual equations (\ref{eq:14}). So, the electric field $\overrightarrow{E}$
and magnetic field $\overrightarrow{B}$ in GDM equations (\ref{eq:38})
are respectively described as the generalized electromagnetic fields
of dyons by combining equations (\ref{eq:5}) and (\ref{eq:13}) in
terms of the components of two four vector potentials $\left\{ A^{\mu}\right\} $
and $\left\{ C^{\mu}\right\} $ \cite{key-15,key-16,key-17} as

\begin{align}
\mathrm{\overrightarrow{\mathrm{E}}} & =-\frac{\partial\overrightarrow{A}}{\partial t}-\overrightarrow{\nabla}\phi-\overrightarrow{\nabla}\times\overrightarrow{C}\nonumber \\
\overrightarrow{\mathrm{B}} & =-\frac{\partial\overrightarrow{C}}{\partial t}-\overrightarrow{\nabla}\varphi+\overrightarrow{\nabla}\times\overrightarrow{A}\label{eq:39}\end{align}
Generalized Dirac Maxwell's (GDM) equations (\ref{eq:38}) are invariant
not only under Lorentz and conformal transformations but are also
invariant under the following duality transformations between electric
$\mathcal{E}$ and magnetic $\mathcal{B}$ quantities i.e.

\begin{align}
\mathcal{E}\Longrightarrow & \mathcal{E}\cos\vartheta+\mathcal{\mathcal{B}\sin\vartheta}\nonumber \\
\mathcal{B}\Longrightarrow & \mathcal{B}\cos\vartheta-\mathcal{\mathcal{E}\sin\vartheta}\label{eq:40}\end{align}
where $\mathcal{E}\equiv\left(\mathsf{e},\overrightarrow{E},\rho,\overrightarrow{j},\phi,\overrightarrow{A}\right)$
and $\mathcal{B\equiv}\left(\mathsf{g},\overrightarrow{B},\varrho,\overrightarrow{k},\varphi,\overrightarrow{C}\right)$.
For a particular value of $\vartheta=\frac{\pi}{2}$, equations (\ref{eq:40})
reduces to

\begin{align}
\mathcal{E}\longmapsto & \mathcal{B}\,\,\,\,\,\,\mathbf{\mathcal{B}}\longmapsto-\mathcal{E}.\label{eq:41}\end{align}
The generalized anti-symmetric dual invariant electromagnetic field
tensors for dyons are written as 

\begin{align}
\mathrm{\mathrm{F^{\mu\nu}}=} & \partial^{\nu}A^{\mu}-\partial^{\mu}A^{\nu}-\frac{1}{2}\varepsilon^{\mu\nu\lambda\sigma}(\partial_{\lambda}C_{\sigma}-\partial_{\sigma}C_{\lambda})=F^{\mu\nu}-\widetilde{\mathcal{F^{\mu\nu}}}\nonumber \\
\widetilde{\mathrm{F^{\mu\nu}}} & =\partial^{\nu}C^{\mu}-\partial^{\mu}C^{\nu}+\frac{1}{2}\varepsilon^{\mu\nu\lambda\sigma}(\partial_{\lambda}A_{\sigma}-\partial_{\sigma}A_{\lambda})=\mathcal{F^{\mu\nu}}+\widetilde{F^{\mu\nu}}.\label{eq:42}\end{align}
The generalized electromagnetic fields of dyons (\ref{eq:39}) are
then be defined as the components of generalized field tensors (\ref{eq:42})
as 

\begin{align}
\mathrm{F}^{0j}= & \mathrm{E}^{j};\,\,\,\,\,\,\widetilde{\mathrm{\mathrm{F}}^{jk}}=\varepsilon^{jkl}\mathrm{B}_{l}\,\,\,(\forall j,k,l=1,2,3)\nonumber \\
\widetilde{\mathrm{F}^{0j}}= & \mathrm{B}^{j};\,\,\,\,\,\,\mathrm{\mathrm{F}}^{jk}=\varepsilon^{jkl}\mathrm{E}_{l}\,\,\,(\forall j,k,l=1,2,3).\label{eq:43}\end{align}
Hence, the covariant form of dual symmetric GDM equations (\ref{eq:38})
is described \cite{key-15,key-16,key-17} as 

\begin{eqnarray}
\mathrm{F}_{\mu\nu,\nu} & = & F_{\mu\nu,\nu}=j_{\mu}\nonumber \\
\mathrm{\tilde{F}}_{\mu\nu,\nu} & = & \mathcal{F}_{\mu\nu,\nu}=k_{\mu}.\label{eq:44}\end{eqnarray}
Therefore, we may write the Lagrangian which follows the minimum action
principle for generalized electromagnetic fields of dyons as 

\begin{align}
\mathcal{L}_{GEM} & =-\frac{1}{4}F_{\mu\nu}F^{\mu\nu}-\frac{1}{4}\mathcal{F}_{\mu\nu}\mathcal{F}^{\mu\nu}+A_{\mu}j^{\mu}+C_{\mu}k^{\mu}.\label{eq:45}\end{align}
This Lagrangian yields the GDM field equations (\ref{eq:44}) and
the Lorentz force equation of motion for dyons as 

\begin{align}
\Im_{\mu} & =F_{\mu\nu}j^{\nu}+\mathrm{\mathcal{F}_{\mu\nu}}k^{\nu}.\label{eq:46}\end{align}
 On the other hand, the dual parts of field tensors giving rise $\partial_{\nu}\widetilde{F^{\mu\nu}}=0$
and $\partial_{\nu}\widetilde{\mathcal{F^{\mu\nu}}}=0$ describe the
Bianchi identities like equation (\ref{eq:37}) for electric and magnetic
charges. If we define the four currents in terms of charge and velocity
as 

\begin{align}
j^{\nu}=\mathsf{e\,}u^{\nu};\,\,\,\,\,\,\,\,\, & k^{\nu}=\mathsf{g}\, u^{\nu},\label{eq:47}\end{align}
 the dual invariant Lorentz force expression (\ref{eq:46}) for dyon
reduces \cite{key-15,key-16,key-17} to 

\begin{align}
\mathsf{\overrightarrow{\Im}=m}\frac{d^{2}\overrightarrow{x}}{dt^{2}} & =e\,(\,\overrightarrow{\mathrm{E}}+\overrightarrow{u}\times\overrightarrow{\mathbf{B}}\,)+\mathsf{g}\,(\,\overrightarrow{\mathrm{B}}-\overrightarrow{u}\times\overrightarrow{\mathrm{E}}\,)\label{eq:48}\end{align}
where $\overrightarrow{u}$ is the velocity of a particle.

\section{$U(1)\times U(1)$ Gauge Formulation of Dyons}

Let us start with the Lagrangian density (\ref{eq:24}) within the
introduction of four spinor Dirac field $\Psi$ for dyons instead
of two component spinor $\psi$ as 

\begin{align}
\Psi & =\left(\begin{array}{c}
\Psi_{1}\\
\Psi_{2}\end{array}\right)\label{eq:49}\end{align}
where $\Psi_{1}$ and $\Psi_{2}$ are two componet spinors. Here $\Psi_{1}$
is identified as the Dirac spinor for a electric charge ( like electron)
while the other spinor $\Psi_{2}$ has been identified as the Dirac
iso-spinors acting on the magnetic monopole. Thus the $\Psi$ may
be visualized as the bi-spinor for dyons in terms of its electric
and magnetic counterparts. Each spinor $\Psi_{1}$and $\Psi_{2}$
satisfy the free particle Dirac equation

\begin{align}
\mathcal{L}_{0}= & \overline{\Psi}(i\gamma^{\mu}\partial_{\mu}+\mathsf{m)\mathrm{\hat{1}}\Psi}\nonumber \\
= & (\overline{\Psi}_{1},\overline{\Psi}_{2})\left[\begin{array}{cc}
(i\gamma^{\mu}\partial_{\mu}+\mathsf{m)} & 0\\
0 & (i\gamma^{\mu}\partial_{\mu}+\mathsf{m)}\end{array}\right]\left(\begin{array}{c}
\Psi_{1}\\
\Psi_{2}\end{array}\right)\nonumber \\
= & \sum_{a=1}^{a=2}\overline{\Psi}_{a}(i\gamma^{\mu}\partial_{\mu}+\mathsf{m)\Psi}_{a}\label{eq:50}\end{align}
where $\hat{1}$ is $2\times2$ unit matrix. So, the Unitary transformations
taking part for the invariance of free particle Dirac equation for
bi-spinor $\Psi$ are the global $U=U^{(e)}(1)\times U^{(m)}(1)$
two component spinors $\Psi_{1}$ and $\Psi_{2}$ . In this case $\Psi_{1}$
acts on unitary gauge group $U^{(e)}(1)$ whereas the iso-spinor $\Psi_{2}$
acts on the other unitary group $U^{(m)}(1)$ with the symbols $(e)$
and $(m)$ are used for the electric and magnetic charges. Thus equation
(\ref{eq:50}) is invariant under global gauge transformation 

\begin{align}
U= & U^{(e)}\times U^{(m)}=\exp(i\Lambda_{j}\tau_{a}^{j}\,^{b})\label{eq:51}\end{align}
where \begin{align}
\tau_{\, a}^{j\,\, b} & =\tau_{\, a}^{1\,\, b}=\left[\begin{array}{cc}
1 & 0\\
0 & 0\end{array}\right]\,\,\,\, and\,\,\,\,\,\tau_{\, a}^{j\,\, b}=\tau^{2}=\left[\begin{array}{cc}
0 & 0\\
0 & 1\end{array}\right]\label{eq:52}\end{align}
and

\begin{align}
\left[\tau_{\, a\,}^{j\,\, b},\tau_{\, a\,}^{k\,\, b}\right] & =\varepsilon_{\,\, l}^{jk}\tau_{\, a\,}^{l\,\, b}=0\label{eq:53}\end{align}
because we have $j,k,l=1,2$. Accordingly the spinor transforms as
\begin{align}
\Psi_{1}\longmapsto\Psi_{1}'\longmapsto\left[U^{(e)}\right]\Psi_{1}= & \exp\{i\Lambda_{1}\}\Psi_{1}\nonumber \\
\Psi_{2}\longmapsto\Psi_{2}'\longmapsto\left[U^{(m)}\right]\Psi_{2}= & \exp\{i\Lambda_{2}\}\Psi_{2}\nonumber \\
\overline{\Psi_{1}}\longmapsto\overline{\Psi_{1}}'\longmapsto\overline{\Psi_{1}}\left[U^{(e)}\right]^{-1}=\overline{\Psi_{1}} & \exp\{i\Lambda_{1}\}\nonumber \\
\overline{\Psi_{2}}\longmapsto\overline{\Psi_{2}}'\longmapsto\overline{\Psi_{2}}\left[U^{(m)}\right]^{-1}=\overline{\Psi_{2}} & \exp\{i\Lambda_{2}\}\nonumber \\
\Psi\longmapsto\Psi^{'}\longmapsto U\Psi & =\left(\begin{array}{cc}
\exp\{i\Lambda_{1}\} & 0\\
0 & \exp\{i\Lambda_{2}\}\end{array}\right)\left(\begin{array}{c}
\Psi_{1}\\
\Psi_{2}\end{array}\right)\nonumber \\
\overline{\Psi}\longmapsto\overline{\Psi}^{'}\longmapsto\overline{\Psi}U^{-1} & =(\overline{\Psi_{1}},\overline{\Psi_{2}})\left(\begin{array}{cc}
\exp\{-i\Lambda_{1}\} & 0\\
0 & \exp\{-i\Lambda_{2}\}\end{array}\right).\label{eq:54}\end{align}
In equations (\ref{eq:51}) and (\ref{eq:53}) $\Lambda_{j}(\forall j=1,2)$
are independent of space and time for global gauge transformations.
If we elevate this symmetry to invariance under local gauge transformations
where $\Lambda_{j}\Longrightarrow\Lambda_{j}(x)\,\,(\forall j=1,2)$
in equations (\ref{eq:51}) and (\ref{eq:53}), the Lagrangian (\ref{eq:50})
transforms as 

\begin{align}
\overline{\Psi}(i\gamma^{\mu}\partial_{\mu}+\mathsf{m)\mathrm{\hat{1}}\Psi} & \longmapsto\overline{\Psi}(i\gamma^{\mu}D_{\mu}+\mathsf{m)\hat{1}\Psi}\label{eq:55}\end{align}
 where partial derivative $\partial_{\mu}$ has been replaced by the
covariant derivative $D_{\mu}$ as

\begin{align}
D_{\mu}=D_{\mu a}^{\,\,\,\, b} & =\partial_{\mu}\delta_{a}^{b}+\beta_{\mu\, j}\tau_{\,\, a}^{j\,\,\, b}\label{eq:56}\end{align}
and $\tau_{\,\, a}^{j\,\,\, b}$ are given by equation (\ref{eq:52})
along with 

\begin{align}
\beta_{\mu\, j}\longmapsto & \beta_{\mu\,1}=A_{\mu}\nonumber \\
\beta_{\mu\, j}\longmapsto & \beta_{\mu\,2}=C_{\mu}.\label{eq:57}\end{align}
These are the gauge potentials respectively associated with the dynamics
of electric and magnetic charges with the following gauge transformations 

\begin{align}
\beta_{\mu\,1}=A_{\mu}\longmapsto A_{\mu}^{\prime}\longmapsto & \left[U^{(e)}\right]A_{\mu}\left[U^{(e)}\right]^{-1}+\frac{1}{\mathsf{e}}\left[U^{(e)}\right]\partial_{\mu}\left[U^{(e)}\right]^{-1}\nonumber \\
\beta_{\mu\,2}=C_{\mu}\longmapsto C_{\mu}^{\prime}\longmapsto & \left[U^{(m)}\right]C_{\mu}\left[U^{(m)}\right]^{-1}+\frac{1}{\mathsf{g}}\left[U^{(m)}\right]\partial_{\mu}\left[U^{(m)}\right]^{-1}\label{eq:58}\end{align}
where 

\begin{align}
\left[U^{(e)}\right] & \Longrightarrow\exp\{i\Lambda_{1}(x)\}\nonumber \\
\left[U^{(m)}\right] & \Longrightarrow\exp\{i\Lambda_{2}(x)\}.\label{eq:59}\end{align}
As such, we may write the $D_{\mu}\Psi$ as 

\begin{align}
D_{\mu}\Psi= & \left[\begin{array}{cc}
\partial_{\mu}-i\mathsf{e}A_{\mu} & 0\\
0 & \partial_{\mu}-i\mathsf{g}C_{\mu}\end{array}\right]\left(\begin{array}{c}
\Psi_{1}\\
\Psi_{2}\end{array}\right)\label{eq:60}\end{align}
which transforms as 

\begin{align}
D_{\mu}\Psi\longmapsto D_{\mu}^{\prime}\Psi^{\prime}\longmapsto & \left(\begin{array}{cc}
\exp\{i\Lambda_{1}(x)\} & 0\\
0 & \exp\{i\Lambda_{2}(x)\}\end{array}\right)\left(\begin{array}{c}
(\partial_{\mu}-i\mathsf{e}A_{\mu})\Psi_{1}\\
(\partial_{\mu}-i\mathsf{g}C_{\mu})\Psi_{2}\end{array}\right)=U(D_{\mu}\Psi).\label{eq:61}\end{align}
 Hence, we get 

\begin{align}
\left[D_{\mu},D_{\nu}\right]\Psi(x) & =\left[\begin{array}{cc}
-i\,\mathsf{e}F_{\mu\nu} & 0\\
0 & -i\,\mathsf{g}\mathcal{F}_{\mu\nu}\end{array}\right]\left(\begin{array}{c}
\Psi_{1}\\
\Psi_{2}\end{array}\right)\label{eq:62}\end{align}
 and it leads to the Jacobi identity 

\begin{align}
\left[D_{\mu},\left[D_{\nu},D_{\lambda}\right]\right]+\left[D_{\nu},\left[D_{\lambda},D_{\mu}\right]\right]+\left[D_{\lambda},\left[D_{\mu},D_{\nu}\right]\right] & =0\label{eq:63}\end{align}
along with the Bianchi identities 

\begin{align}
D_{\mu}F_{\nu\lambda}+D_{\nu} & F_{\lambda\mu}+D_{\lambda}F_{\mu\nu}=0\nonumber \\
D_{\mu}\mathcal{F}_{\nu\lambda}+D_{\nu} & \mathcal{F}_{\lambda\mu}+D_{\lambda}\mathcal{F}_{\mu\nu}=0\label{eq:64}\end{align}
As such, the total Lagrangian for generalized fields of dyons is described
as

\begin{align}
\mathcal{L}=-\frac{1}{4}F_{\mu\nu}F^{\mu\nu}-\frac{1}{4} & \mathcal{F}_{\mu\nu}\mathcal{F^{\mu\nu}}+\overline{\psi}(i\gamma^{\mu}\partial_{\mu}+\mathsf{m})\psi-A_{\mu}j^{\mu}-C_{\mu}k^{\mu}\label{eq:65}\end{align}
where \begin{align}
j^{\mu}=\mathsf{e} & \overline{\Psi_{1}}\gamma^{\mu}\Psi_{1}\label{eq:66}\end{align}
and \begin{align}
k^{\mu}=\mathsf{g} & \overline{\Psi_{2}}\gamma^{\mu}\Psi_{2}\label{eq:67}\end{align}
are the the four currents associated respectively with electric and
magnetic charges on dyons. These four-currents obtained from the Dirac
spinor $\Psi_{1}$ and the Dirac iso-spinor $\Psi_{2}$ satisfy the
following conserved relations

\begin{align}
\partial_{\mu\,}j^{\mu} & =j^{\mu},\mu=0\label{eq:68}\end{align}
and

\begin{align}
\partial_{\mu\,}k^{\mu} & =k^{\mu},\mu=0.\label{eq:69}\end{align}
Like other electric and magnetic dynamical variables, one can introduce
duality transformations between the electric and magnetic gauges corresponding
to orthogonal transformations in group space i.e. 

\begin{align}
\Lambda_{1}\Longrightarrow & \Lambda_{1}\cos\vartheta+\Lambda_{2}\mathcal{\sin\vartheta}\nonumber \\
\Lambda_{2}\Longrightarrow & \Lambda_{2}\cos\vartheta-\Lambda_{1}\mathcal{\sin\vartheta}\label{eq:70}\end{align}
and with the use of constancy condition\cite{key-16}

\begin{align}
\frac{\mathsf{g}}{\mathsf{e}} & =\frac{C_{\mu}}{A_{\mu}}=\frac{k_{\mu}}{j_{\mu}}=\frac{\Lambda_{2}}{\Lambda_{1}}=\frac{\mathcal{F}_{\mu\nu}}{F_{\mu\nu}}=\frac{\widetilde{\mathrm{F^{\mu\nu}}}}{\mathrm{F}^{\mu\nu}}=-tan\,\theta=Constant\label{eq:71}\end{align}
we get

\begin{align}
\left[D_{\mu},D_{\nu}\right]\Psi(x) & =\left[\begin{array}{cc}
-i\,\mathsf{e}\mathrm{F_{\mu\nu}} & 0\\
0 & -i\,\mathsf{g}\widetilde{\mathcal{\mathrm{F}}}_{\mu\nu}\end{array}\right]\left(\begin{array}{c}
\Psi_{1}\\
\Psi_{2}\end{array}\right)\label{eq:72}\end{align}
where $\mathrm{F_{\mu\nu}}$ and $\widetilde{\mathcal{\mathrm{F}}}_{\mu\nu}$
are the generalized dual invariant electromagnetic fields of dyons
and satisfy independently the Bianchi identity (\ref{eq:64}). Hence,
the Lagrangian density (\ref{eq:65}) may now be written as

\begin{align}
\mathcal{L}=-\frac{1}{4}\mathrm{F}_{\mu\nu}\mathrm{F}^{\mu\nu}-\frac{1}{4} & \widetilde{\mathrm{F}}_{\mu\nu}\mathrm{\widetilde{F}}^{\mu\nu}+\overline{\psi}(i\gamma^{\mu}\partial_{\mu}+\mathsf{m})\psi-A_{\mu}j^{\mu}-C_{\mu}k^{\mu}\label{eq:73}\end{align}
where $\mathrm{F_{\mu\nu}}$ and $\widetilde{\mathcal{\mathrm{F}}}_{\mu\nu}$
transforms as 

\begin{align}
U\left[D_{\mu},D_{\nu}\right]U^{-1}\Psi(x) & \Rightarrow\left[\begin{array}{cc}
-i\,\mathsf{e}\left[U^{(e)}\right]\mathrm{F_{\mu\nu}}\left[U^{(e)}\right]^{-1} & 0\\
0 & -i\,\mathsf{g}\left[U^{(m)}\right]\widetilde{\mathcal{\mathrm{F}}}_{\mu\nu}\left[U^{(m)}\right]^{-1}\end{array}\right]\left(\begin{array}{c}
\Psi_{1}\\
\Psi_{2}\end{array}\right)\nonumber \\
\Longrightarrow\mathrm{F_{\mu\nu}}\longmapsto & \left[U^{(e)}\right]^{-1}\mathrm{F_{\mu\nu}}\left[U^{(e)}\right]\nonumber \\
\Longrightarrow\mathrm{\widetilde{F}_{\mu\nu}}\longmapsto & \left[U^{(m)}\right]^{-1}\mathrm{\widetilde{F}_{\mu\nu}}\left[U^{(m)}\right].\label{eq:74}\end{align}
So, it is observed that the Lagrangian density (\ref{eq:73}) reproduces
the dual symmetric and Lorentz covariant generalized Dirac Maxwell's
(GDM) field equations (\ref{eq:44}) and Lorentz force equation (\ref{eq:46})
of motion for two potential theory of dyons. Thus, with the use of
Jacobi identity (\ref{eq:63}), we get the Bianchi identity for generalized
electromagnetic field tensors $\mathrm{F_{\mu\nu}}$ and $\widetilde{\mathcal{\mathrm{F}}}_{\mu\nu}$
as 

\begin{align}
D_{\mu}\mathrm{F}_{\nu\lambda}+D_{\nu} & \mathrm{F}_{\lambda\mu}+D_{\lambda}\mathrm{F}_{\mu\nu}=0\nonumber \\
D_{\mu}\mathcal{\widetilde{\mathrm{F}}}_{\nu\lambda}+D_{\nu} & \widetilde{\mathcal{\mathrm{F}}}_{\lambda\mu}+D_{\lambda}\mathcal{\mathrm{\widetilde{F}}}_{\mu\nu}=0.\label{eq:75}\end{align}
As such, the classical theory of dyons has been verified and the incorporation
of two four potentials in generalized electromagnetic fields of dyons
has been justified in the frame work of $U(1)\times U(1)$ gauge theory
where the fist unitary Abelian gauge group $U^{(e)}(1)$ acts on the
Dirac spinors due to the presence of the electric charge while second
unitary Abelian gauge group $U^{(m)}(1)$ acts on the Dirac iso-spinors
due to the presence of the magnetic charge on dyons. The activation
of gauge group $U^{(m)}(1)$ on Dirac iso-spinor is advantageous so
that it may further be extended to enlarge the gauge group to describe
the non-Abelian correspondence of monopoles (dyons) in current grand
unified and supersymmetric gauge theories associated with dyons. Consequently,
$U(1)\times U(1)$ gauge group may further be extended in order to
describe the built in duality between the $SU(2)\times U(1)$ gauge
theory of electro-weak interaction and $SU(2)\times U(1)$ gauge theory
of gravity described earlier by Dehnen et al \cite{key-18} so that
one can be able to understand better the current grand unified theories,
supersymmetry, super-gravity and super-strings.
\begin{description}
\item [{Acknowledgment-}]~
\end{description}
One of us OPSN is thankful to German Academic Exchange Service (Deutscher
Akademischer Austausch Dienst), Bonn for their financial support under
DAAD re-invitation programme at Universität Konstanz.

\end{document}